\documentclass[entropy,preprints,accept,moreauthors,pdftex]{Definitions/mdpi} 

\firstpage{1} 
\pubvolume{1}
\issuenum{1}
\articlenumber{0}
\pubyear{2021}
\copyrightyear{2021}
\externaleditor{Academic Editor: Ryszard Kutner} 
\datereceived{} 
\dateaccepted{} 
\datepublished{} 
\hreflink{https://doi.org/}

\Title{Financial Return Distributions: Past, Present, and COVID-19}

\TitleCitation{Financial Return Distributions: Past, Present, and COVID-19}

\Author{Marcin Wątorek $^{1}$\orcidA{}, Jarosław Kwapień $^{2,}$\orcidC{} and Stanisław Drożdż $^{1,2}$\orcidB{}}

\AuthorNames{Marcin Wątorek, Jarosław Kwapień and Stanisław Drożdż}

\AuthorCitation{Wątorek, M.; Kwapień, J.; Drożdż, S.}

\address{
$^{1}$ \quad Faculty of Computer Science and Telecommunications, Cracow University of Technology, ul.~Warszawska 24, 31-155 Krak\'ow, Poland; marcin.watorek@pk.edu.pl (M.W.); Stanislaw.Drozdz@ifj.edu.pl (S.D.) \\
$^{2}$ \quad Complex Systems Theory Department, Institute of Nuclear Physics, Polish Academy of Sciences, ul.~Radzikowskiego 152, 31-342 Krak\'ow, Poland
}

\corres{Correspondence: jaroslaw.kwapien@ifj.edu.pl}

\abstract{We analyze the price return distributions of currency exchange rates, cryptocurrencies, and contracts for differences (CFDs) representing stock indices, stock shares, and commodities. Based on recent data from the years 2017--2020, we model tails of the return distributions at different time scales by using power-law, stretched exponential, and $q$-Gaussian functions. We focus on the fitted function parameters and how they change over the years by comparing our results with those from earlier studies and find that, on the time horizons of up to a few minutes, the so-called ``inverse-cubic power-law'' still constitutes an appropriate global reference. However, we no longer observe the hypothesized universal constant acceleration of the market time flow that was manifested before in an ever faster convergence of empirical return distributions towards the normal distribution. Our results do not exclude such a scenario but, rather, suggest that some other short-term processes related to a current market situation alter market dynamics and may mask this scenario. Real market dynamics is associated with a continuous alternation of different regimes with different statistical properties. An example is the COVID-19 pandemic outburst, which had an enormous yet short-time impact on financial markets. We also point out that two factors---speed of the market time flow and the asset cross-correlation magnitude---while related (the larger the speed, the larger the cross-correlations on a given time scale), act in opposite directions with regard to the return distribution tails, which can affect the expected distribution convergence to the normal distribution.}

\keyword{return distributions; power-law tails; stretched exponentials; $q$-Gaussians; financial markets; COVID-19}

\begin{document}
\section{Introduction}

A proper risk assessment is one of the key prerequisites of any prospective financial investment. Even for an asset of moderate volatility, underestimating the probability of occurrence of an event of a given magnitude can lead to severe outcomes. Among the methods of dealing with risk assessment is the determination of a correct probability distribution function for asset price fluctuations in order to construct an adequate model of that asset's price dynamics. This issue has been of central interest since the early years of econometrics. It was Bachelier who proposed a model of the stock option price dynamics based on an uncorrelated random walk with a Gaussian distribution of fluctuations~\cite{bachelier1900}. Later, it was found that the Gaussian noise hypothesis was only a poor approximation of the empirical data, which shows non-vanishing higher moments of the fluctuation distributions, i.e., skewness and positive excess kurtosis. Based on an observation of the cotton price dynamics, Mandelbrot proposed to model the logarithmic price increments (returns) with a process of L\'evy flights, which is described by a heavy-tailed probability distribution function that is stable~\cite{levy1925,paul1999}. These distributions are defined by their characteristic function as they do not have a closed analytic form. However, their tails decrease as a power law in the limit of large $x$:
\begin{equation}
L_{\alpha}(x) \sim {1 \over |x|^{1+\alpha}}, \quad x \to \pm \infty,
\end{equation}
where $0 < \alpha < 2$.
 
According to Mandelbrot~\cite{mandelbrot1963}, such a process can account for the absence of a convergence of the aggregated return distribution to the normal distribution as expected by the central limit theorem (CLT). The heavy tails are thus viewed as a natural limit of the aggregated independent or weakly dependent factors provided they are described by the stable distributions. However, this hypothesis has a weak point because the empirical data cannot exhibit the infinite variance required to maintain the distribution stability under aggregation. After the pioneering work of Mandelbrot, many researchers investigated financial time series in order to verify his outcomes. For example, Fama reported that the daily returns of stocks are better approximated by the infinite-variance distribution than the normal distribution or a mixture of the normal distributions~\cite{fama1965}. The L\'evy stability of the return pdfs in their central parts was also confirmed, among others, by Blume ($\alpha\approx 1.7-1.8$)~\cite{blume1970}, Teichmoeller ($\alpha\approx1.6$)~\cite{teichmoeller1970}, and Blattberg and Gonedes ($\alpha\approx 1.6$)~\cite{blattberg1974}. Some reports pointed out that, although central parts of the return distribution can be approximated by the stable distributions, the same cannot be said about the distant parts of their tails, which decay faster than expected. Officer found that the tails of the daily and monthly return distributions are no doubt thicker than Gaussian but at the same time thinner than L\'evy-stable~\cite{officer1972}. Barnea et al. observed that the daily return distributions for some stocks are well approximated by stable distributions, while for other stocks, they are not~\cite{barnea1973}. Much later, Young and Graff reported that the real-estate annual return distribution can be fitted by a stable function using $\alpha\approx 1.5$~\cite{young1995}.

Along with the research on empirical data, much effort was devoted to developing models that could mimic the market dynamics. Among such models, the subordinated stochastic processes do not require an assumption of the L\'evy-stable character of the underlying dynamics and assume that the price movement is a Brownian motion that takes place in time, which itself is a stochastic process with positive increments and finite variance (e.g., a lognormal process)~\cite{clark1973}. In practice, the subordinating process is assumed to be volume or transaction number. As an alternative, Engle proposed that the distribution tails are heavy because of the heteroskedasticity of the return-generating process, in which large returns are caused by a locally large variance of the process~\cite{engle1982}. Mantegna and Stanley found a dual structure of the stock index return distribution (S\&P500 index during the years 1984--1989), with its central part being in agreement with a L\'evy-stable distribution and with exponentially decaying distant tails~\cite{mantegna1995}:
\begin{equation}
L_{\alpha,\gamma}^{\rm tr}(x) \sim {1 \over |x|^{1+\alpha}} e^{-\gamma|x|}, \quad \gamma > 0.
\end{equation}

While considering the aggregated returns at different time horizons, they did not find any trace of a convergence to the normal distribution. Based on these findings, they proposed a new model for the price return dynamics: a truncated L\'evy flight process. They also showed that the heteroskedastic model (GARCH) does not fit the data well~\cite{mantegna1995}.
This type of distribution ($\alpha\approx$ 1.6--1.7) was also reported from an analysis of the same S\&P500 index recorded {over} a longer interval (1986--2000). In contrast, the aggregated returns showed a crossover to a CLT regime around a time scale of 20 days~\cite{miranda2003}.

Plerou et al. and Gopikrishnan et al. presented two parallel, comprehensive studies of the stock market high-frequency data representing stock price returns for 1000 American companies and S\&P500 index returns~\cite{plerou1999,gopikrishnan1999}, in which they observed the cumulative distribution function tails obtained from aggregated returns over a substantial spectrum of time scales from 5 min (stocks) and 1 min (index) to 4 years. They found that the return distributions have power law tails, with the exponent $2.5 < \alpha < 4$ depending on a stock. However, despite the fact that they did not fit the L\'evy-stable domain ($\alpha < 2$), these distributions were invariant under a change in the time scale up to $\Delta t=16$ days. Only for the sampling intervals longer than 16 days, a slow transition to a normal distribution was observed~\cite{plerou1999}. An analogous invariance of the return distribution shape with the power exponent $\alpha\approx 3$ under the time-scale change was observed for the S\&P500 index, but in that case, the crossover occurred earlier at $\Delta t=4$ days. Only for the time scales longer than 4 days, a slow convergence to a Gaussian distribution was seen. A similar behavior was found in the indices from other stock markets (Nikkei \& Hang-Seng)~\cite{gopikrishnan1999}. This surprising behavior of the stock markets led the authors to formulate the so-called ``inverse cubic law''---a conjecture that the power-law tails of the return distributions with the scaling exponent $\alpha\approx 3$ are a universal property of all stock markets at short and medium time scales~\cite{gopikrishnan1998}. Indeed, similar statistical characteristics were found by other researchers in data collected from other stock markets~\cite{lux1996,%
makowiec2001,drozdz2001a,lillo2002,kaizoji2002,drozdz2003,kim2003,%
kim2004,coronel-brizio2005,sinha2006,oh2006,rak2007,gu2007,%
drozdz2007,wang2009,kwapien2012,eom2019}, Forex~\cite{watorek2019}, commodity markets~\cite{matia2002,watorek2019}, and the cryptocurrency market~\cite{drozdz2018,watorek2019,%
watorek2021,takaishi2021}.

The only possible explanation of this result is that the analyzed data violated the assumptions of the central limit theorem, i.e., the returns were significantly correlated. Indeed, the cross-correlations among the stock returns representing different companies are an obvious characteristics of all stock markets~\cite{epps1979,plerou2002,kwapien2003,kwapien2004,kwapien2006,kwapien2012}. It was shown that the inter-stock cross-correlation strength has a strong impact on the index return distributions and can even modify their tail behavior, leading to a kind of alternation between different power-law regimes: stable and unstable~\cite{kwapien2003}. On the other hand, the cross-correlations between different stock markets can also induce a significant regime change~\cite{drozdz2001a,lillo2002}. The existence of  autocorrelation in returns is a more delicate issue: while the returns reveal some short-term memory lasting for a few minutes, the existence of long-term memory is doubtful~\cite{lo1991,mills1993,plerou1999,gopikrishnan1999,drozdz2003}, even though there were reports stating that the returns can show some autocorrelation or persistence over long terms~\cite{fama1988,wright2001,henry2002,podobnik2006,hull2014}. On the other hand, there is consensus over a fact that the long-term autocorrelation is present in absolute returns (volatility) and in some more fundamental observables such as fluctuations in stock market orders, transaction size, and market liquidity~\cite{ding1993,baillie1996,lillo2004}. The existence of a return autocorrelation can be considered an important factor that can destroy market efficiency~\cite{fama1970,alvarez-ramirez2012}. These ubiquitous manifestations of the inverse cubic scaling in the financial data encouraged Gabaix et al. to propose a model that was able to account for this phenomenon~\cite{gabaix2003}. According to this model, the inverse-cubic return fluctuations were a result of two processes: the volume fluctuation that forms a probability distribution function with the tail index 1.5 and a specific square-root form of the price impact function, which together produce a tail index equal to 3~\cite{gabaix2003}. However, 
Farmer and Lillo pointed out that the price impact function is specific to individual markets and even to individual stocks; thus, it cannot produce any universal behavior. Also, the dependence on transactions is slower than the square root and the volumes are not power-law distributed, so they cannot lead to a power-law behavior of the returns with $\alpha\approx 3$~\cite{farmer2004}. The price changes are driven by more factors than simply volume and transaction number fluctuations---it can be the order book structure, for example~\cite{gillemot2006,taranto2018}. Moreover, there is plenty of published evidence that various financial assets either do not have the power-law distribution tails~\cite{kaizoji2003,silva2004,malevergne2005,malevergne2006,oh2006,cortines2007,ren2009} or their scaling exponent $\alpha$ differs from 3 even for the short time scales~\cite{mart2004,yang2006,scalas2007,suarez-garcia2013,rak2013,begusic2018,watorek2019}. Given these results together, the inverse cubic scaling cannot be considered a universal property of financial returns and, thus, cannot be called ``a law''. However, it manifests itself sufficiently often to allow us to view it as one of the possible reference models describing the empirical return distribution tails {(there is a plethora of volatility models, which takes into account various factors; a review of such models can be found in Poon and Granger~\cite{poon2003})}.

The power law tails of the return distributions, which are among the financial stylized facts, can be reproduced with a broad range of the scaling exponent by means of various models based on stochastic \mbox{processes~\cite{solomon1998,solomon2001,lux2002,sornette2001,sornette2002,dragulescu2002,silva2004,alejandro-quinones2006,bormetti2008,gerig2009}}, including multiplicative \mbox{processes~\cite{ghashghaie1996,mandelbrot1997,breymann2000}}, the minority game and other agent-related dynamics~\cite{bak1996,caldarelli1997,cont2000,zhang2000,challet2001}, as well as spin dynamics~\cite{bornholdt2001,kaizoji2002}.

Apart from power-law functions, the tail behavior of the return distributions can also be approximated by exponential functions and stretched exponential functions~\cite{laherrere1998}. The latter are defined by the following expression:
\begin{equation}
f(x) \sim \exp{x^{-\beta}}, \qquad 0 < \beta < 1.
\end{equation}

Such a functional form allows for the stretched exponents to locally resemble the power laws. There were many published studies in which the return distributions were approximated successfully by the exponents, and some researchers advocate using these functions instead of the power laws~\cite{kaizoji2002,dragulescu2002,silva2004,matia2004,malevergne2005,malevergne2006,pisarenko2006,yang2006,oh2006,gu2007,cortines2007,ren2009}. Another type of exponential function that is sometimes considered in the context of financial data is the Laplace distribution function {$p(x) \sim \exp(-|x|)$}. This function can also demonstrate heavy tails. It was observed that some empirical return distributions can be approximated by this function~\cite{linden2001,kaizoji2003}.

The functions that have been discussed so far do not exhaust the possible models that can be used to approximate the empirical return distributions. In a financial context, a  particularly important class is the $q$-Gaussian functions. They were derived as a part of the formalism of nonextensive statistical mechanics based on the Tsallis nonadditive entropy~\cite{tsallis1988}:
\begin{equation}
S_q = k_{\rm B} {1 - \int [p(x)]^q dx \over q-1},
\end{equation}
where $p(x)$ is some probability distribution and $k_{\rm B}$ is a positive constant. Under certain conditions, this entropy is maximized by a family of $q$-Gaussian distributions given by
\begin{equation}
G_q(x) \sim \exp_q[-\mathcal{B}_q(x-\mu_q)^2],
\end{equation}
where
\begin{equation}
\exp_q x=[1+(1-q_x]^{1 \over 1-q}, \quad \mathcal{B}_q = [(3-q)\sigma_q^2]^{-1},
\label{eq::q.exponent}
\end{equation}
provided that $0 < q < 3$ and that $\mu_q$ and $\sigma_q^2$ are \mbox{$q$-mean} and $q$-variance, respectively. The \mbox{$q$-Gaussians} generalize both the normal distribution ($q=1$) and the L\'evy distributions ({$5/3 < q < 3$}). Their attractiveness comes from the fact that, for the correlated random variables, the \mbox{$q$-Gaussians} become stable distributions. Moreover, their tail behavior can also resemble the power laws~\cite{tsallis2009}. As the price returns are correlated, one can expect that these functions can describe the statistical properties of returns. Indeed, there is a growing evidence that the $q$-Gaussian distributions can approximate the empirical return \mbox{distributions~\cite{michael2003,cortines2007,rak2007,drozdz2007,mu2010,drozdz2009}.}

The $q$-Gaussians are among the functions borrowed from the nonextensive statistical mechanics that were exploited in this context. Another example is the $q$-exponent given by Equation (\ref{eq::q.exponent}), which was also reported to fit the empirical returns from a stock market~\cite{queiros2004}. Finally, some researchers consider the normal-inverse Gaussian function to be a prospective model that can successfully be fitted to the data~\cite{suarez-garcia2013}.

This short review of the return distribution modeling approaches shows that there is a cornucopia of the reported results that were even contradictory sometimes. The only firm observation that is shared by all the studies is that the return distributions reveal heavy tails, at least at short time scales. On medium and long time scales, the situation depends strongly on a data set, a market, and a financial instrument. Dro\.zd\.z et al. attempted to resolve this problem by noticing that the most well-known results regarding the return distributions, i.e.,  Mandelbrot's L\'evy stability ($\alpha < 2$)~\cite{mandelbrot1963};  Mantegna and Stanley's truncated L\'evy flights~\cite{mantegna1995}; Plerou and Gopikrishnan's unstable power-law tails ($\alpha\approx 3$), which are persistent under aggregation of the returns until the time scales of days or even a month~\cite{plerou1999,gopikrishnan1999}; and their own results with the $\alpha\approx 3$ regime already breaking at the time scale of hours~\cite{drozdz2003}, were based on the data covering different epochs: 1816--1958 (Mandelbrot), 1984--1989 (Mantegna), 1926--1995 (Plerou and Gopikrishnan), and 1998--1999 (Dro\.zd\.z). One can {follow} the whole historical process of the financial market development, introduction of new financial instruments, technological innovations, transition from the classic ``floor-based'' markets to the digital markets, computing power increase, telecommunication revolution, etc. from past to present. This inevitably leads to the constantly increasing number of investors, transactions, and pieces of information that arrive at the market. These are accompanied by the increasing amount of money and information processing speed, which, if taken together, result in an overall acceleration of the market time flow. Any unit of time nowadays corresponds to a much longer interval in the past. From this perspective, the market properties once {observed, say, at a daily scale}, now can be observed at scales of hours, minutes, or even seconds. This may be the very reason why Mandelbrot observed the L\'evy-stable distributions that are hardly seen today and why Plerou and Gopikrishnan reported the crossover to the CLT-related convergence of distribution tails at the time scale of many days, while today, such a behavior is observed within hours or minutes. This hypothesis formulated by Dro\.zd\.z et al. was later supported by other analyses as well~\cite{drozdz2003,yang2006,rak2007,drozdz2007,rak2013}.

However, based on data covering a given time interval, one can observe an analogous phenomenon by considering, e.g., the stocks representing companies with different capitalization. Since there is statistically a relation between the capitalization of a company and the number of transactions involving its stock shares, the highly capitalized stocks ``feel'' that time flows faster than their lower-capitalized counterparts. In consequence, the properties of the corresponding return distributions substantially differ between both groups, with the former displaying thicker tails than the latter~\cite{lo1988,drozdz2003,mu2010,kwapien2012,eom2019}. Qualitatively similar observation can be made by comparing the distributions for the data from the markets of different developmental stage, e.g., the mature markets and the emerging markets. The former are characterized by higher liquidity and a higher transaction number than the latter; therefore, generally, the situation is parallel to the previous cases. Studies of the data from the emerging markets report thick tails with small scaling exponents more frequently than the mature markets~\cite{bekaert1998,kim2003,kim2004,matia2004,lee2004,sarkar2006,vicente2006,sinha2006,cortines2007,alfonso2012,gang2012,gang2013,hull2014}.

Another issue related to return distributions is their asymmetry between positive and negative parts. It was investigated in various works as it is also an important factor in investment risk assessments (the gain--loss asymmetry). Typically, this property was tested by means of the third moment (skewness) of return distributions, in which a negative value means a higher probability of a significant gain with respect to a significant loss while a positive value means the opposite. The negative skewness is associated, thus, with a positive tail of the distribution being heavier than the negative tail. There are mixed outcomes of the empirical data reported in the literature, including indications of either positive, negative, or neutral skewness as well as the scaling exponent difference between the left and the right tails (in the case of power-law tails) dependent on the analyzed time intervals, markets, and securities (e.g.,~References \cite{samuelson1970,kane1977,friend1980,kon1984,mueller1990,guillaume1997,mantegna1995,gopikrishnan1999,plerou1999,makowiec2001,kaizoji2003,matia2004,sinha2006,gu2007,coronel-brizio2007,suarez-garcia2013,watorek2019,derksen2020,miskiewicz2021}). However, {even though} a difference between the positive and negative tails exists in the data{,} it has a much weaker impact on the distribution shape and the related investment risk than the heavy tails. Therefore, in many studies reported in the literature, only absolute returns are considered, neglecting their actual signs (e.g.,~References \cite{plerou1999,gopikrishnan1999,drozdz2003,drozdz2007,rak2007,watorek2021}). As our study is focused on an investigation of the tail exponent stability with respect to the time scale $\Delta t$ and, based on literature and our previous experience, we expect larger effects due to the time-scale change than due to the left--right tail asymmetry, we neglect the return sign and consider both tails together by analyzing the absolute return values. { In fact, our major new finding is that, in recent years, the market's ``internal'' time stopped accelerating with respect to our ordinary ``clock'' time. Other factors also affect the convergence of return distributions to the Gaussian with increasing $\Delta t$, especially those that cause extreme volatility and strong cross-correlations between assets such as COVID-19. We discuss the interplay of these two factors in the following sections.}

The remainder of our paper is organized as follows: in Section~\ref{sect::data}, we present the data sets that were analyzed; in Section~\ref{sec3}, we discuss the results; and in Section~\ref{sec4}, we collect the main conclusions of our study.

\section{Data}
\label{sect::data}

We analyzed recent tick-by-tick recordings of the contracts for differences (CFDs) representing (1) six major stock market indices, CAC40 (Euronext), DAX30 (Deutsche B\"orse), FTSE100 (London SE), DJIA (New York SE), S\&P500 (New York SE \& NASDAQ), and NASDAQ100 (NASDAQ); (2) 240 U.S. stock shares and 30 stock shares with the highest capitalization from Germany, France, and the U.K. (see Appendix~\ref{sect::app} for their list); (3) four commodities, U.S. crude oil (CL), high grade copper (HG), silver (XAG), and gold (XAU); (4) the currency exchange rates (not CFDs) involving five major currencies, USD, EUR, GBP, CHF, and JPY; and (e) two cryptocurrencies, bitcoin (BTC)~\cite{nakamoto2008} and ethereum (ETH)~\cite{ethereum}.  The commodity CFD prices are expressed in U.S. dollars. The data comes from Dukascopy (the index, stock share, commodity CFDs, and currency exchange rates)~\cite{dukascopy} and Kraken exchange {(cryptocurrencies)}~\cite{kraken} and covers 4 years from January 2017 to December 2020 (except for the stock share CFDs that cover a shorter interval starting from January 2018). Different instrument types have different trading hours, with the stock market index and commodity CFDs quoted from Monday to Friday (00:00--23:00 hours CET, daylight saving time-adjusted), the stock share CFDs quoted from Monday to Friday (U.S.: 15:30--22:00 CET, European: 09:00--17:30 CET), the currency exchange rates quoted around the clock from Monday to Friday, and the cryptocurrency exchange rates quoted continuously 24/7.

\begin{figure}[H]
\includegraphics[width=0.7\textwidth]{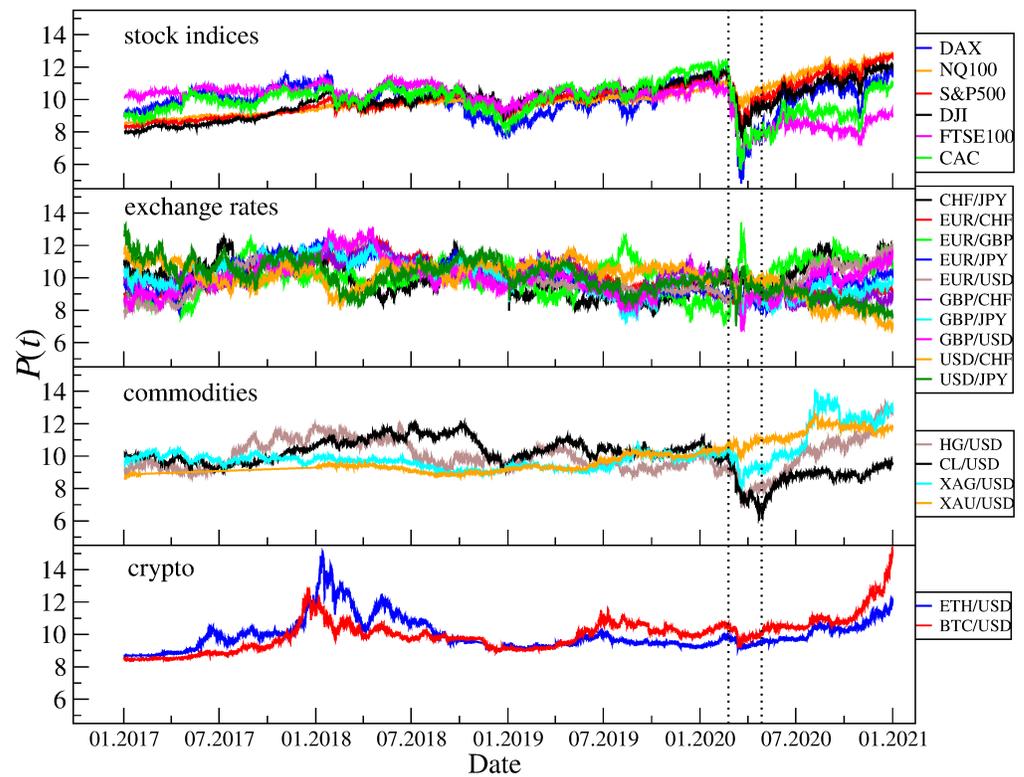}
\caption{Evolution of the CFD price and exchange rate quotations of various assets over the 4~year interval 2017--2020 (data source: Dukascopy~\cite{dukascopy}) and the cryptocurrency prices (data source: Kraken~\cite{kraken}). The quotations have been standardized in order to facilitate comparison. The vertical dashed lines indicate the COVID-19 outburst in March--April 2020.}
\label{fig::quotations}
\end{figure}

\begin{figure}[H]
\includegraphics[width=0.7\textwidth]{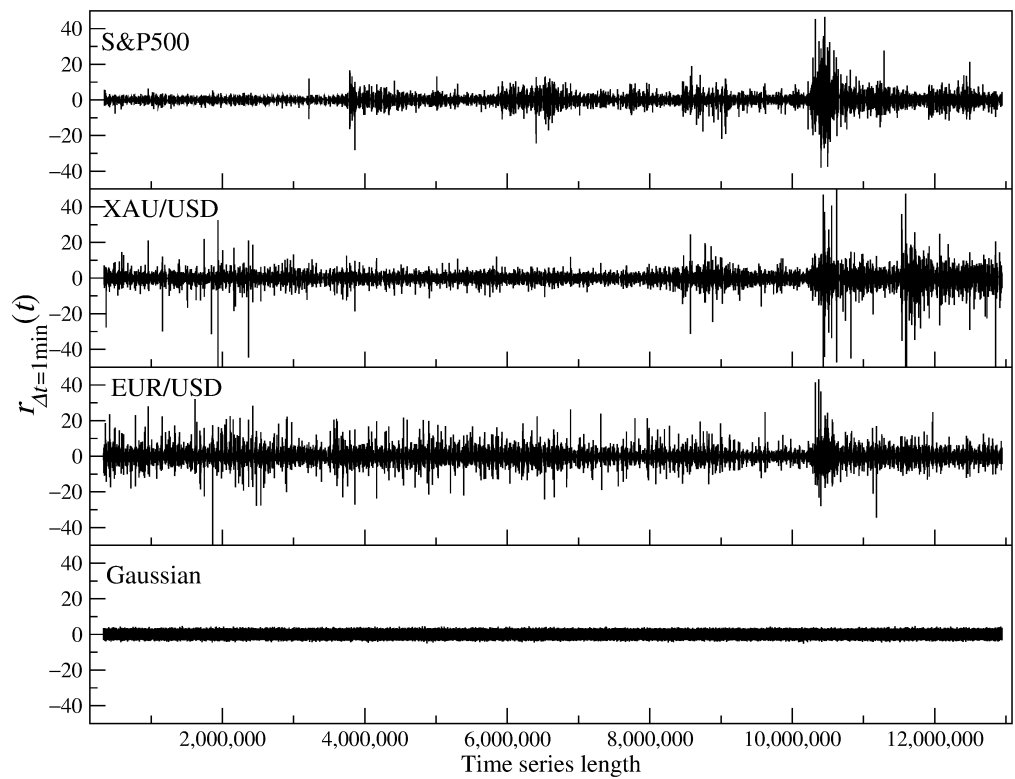}
\caption{Time series of the standardized 1-min returns of sample financial instruments, S\&P500 CFDs, gold CFDs (XAU/USD), and EUR/USD, together with a Gaussian noise of the same length. Note the leptokurtic character of the empirical data.}
\label{fig::returns}
\end{figure}

Price $P(t)$ of an asset is defined at the moment of transaction only and remains undefined otherwise. Therefore, in order to construct an evenly sampled time series of the price quotations, we assume that the price remains constant between the consecutive transactions, which is standard practice. The quotations of all the instruments were sampled with $\Delta t$= 1 s, 10 s, 1 min, 10 min, and 1 h frequency and transformed into the normalized logarithmic returns $r_{\Delta t}$ according to
\begin{equation}
r_{\Delta t}=(R_{\Delta t}-\mu_R)/ \sigma_R, \quad R_{\Delta t}(t)=\log(P(t+\Delta t))-\log(P(t)),
\end{equation}
where $\mu_R$ and $\sigma_R$ are the mean and standard deviation of $R_{\Delta t}(t)$, respectively, and $\Delta t$ is a sampling interval. For each asset, we obtained five time series representing the returns for different time scales $\Delta t$. Figure~\ref{fig::quotations} shows the evolution of $P(t)$ for various assets that are analyzed in our work. The COVID-19 outburst in the U.S. in March and April 2020 that had a strong impact on all financial markets has been distinguished by vertical lines. A few corresponding time series of the normalized returns $r_{\Delta t}(t)$ with $\Delta t$=1 min are shown in Figure~\ref{fig::returns} together with a simulated Gaussian noise of the same length.


\section{Results}\label{sec3}

\begin{figure}[H]

\includegraphics[width=0.7\textwidth]{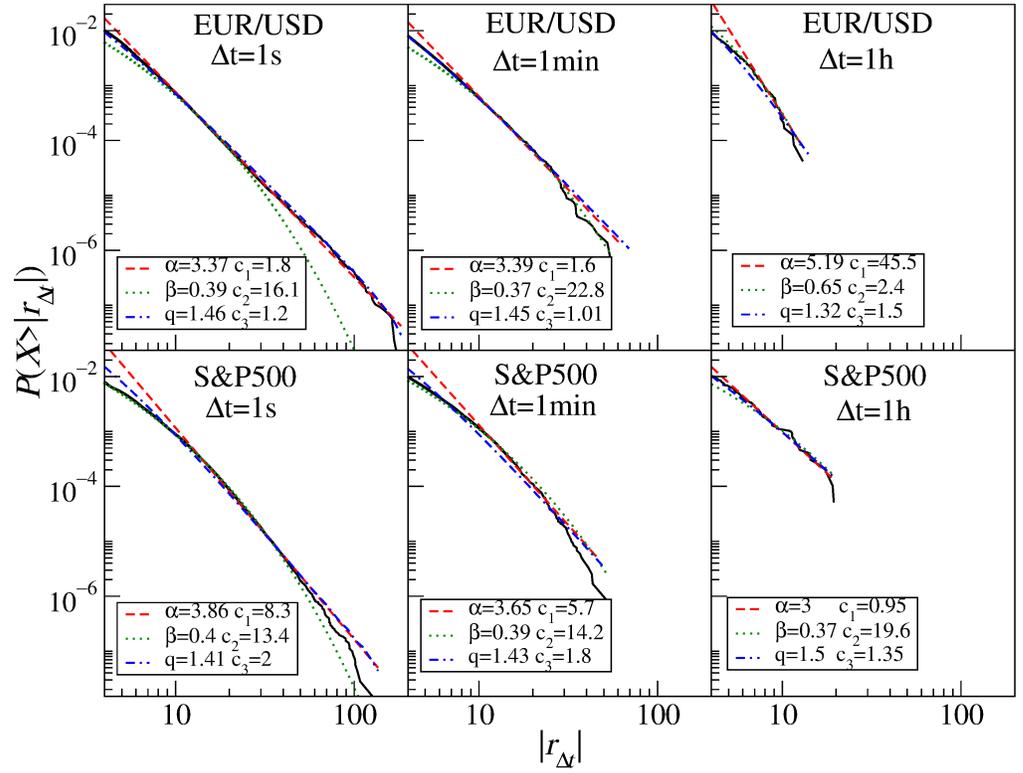}
\caption{The least-square best fits of the power-law function (red dashed), the stretched exponential (green dotted), and the $q$-Gaussian function (blue dash-dotted). Sample cumulative distribution functions of the returns for the EUR/USD exchange rate (top) and the S\&P500 index CFDs (bottom) are shown with different sampling intervals $\Delta t$ from 1 s to 1 h. }
\label{fig::best_fits}
\end{figure}

For each individual time series of the absolute normalized returns, we created a cumulative distribution function and investigated how fast its tail decays. In order to quantify the tail behavior, we fit the empirical histograms with selected models that are of the highest significance in this context: the power-law function, the stretched exponential function, and the $q$-Gaussian function. Figure~\ref{fig::best_fits} shows sample return distributions with the three best-fitted models of interest. We refer the reader to the specific subsections for a discussion on the asset statistical properties; here, we consider only the fits. In each panel, it is evident that the power-law function (dashed line) is able to reproduce the empirical histograms in their far-tail region while it fails to describe the central part of the distributions completely. The stretched exponential and $q$-Gaussian functions perform much better in the central parts, while only the latter works well in the tails. However, as the $q$-Gaussian and power-law functions converge to the same behavior in the tail regions and as both the parameters $\alpha$ and $q$ are related with each other via a relation,
\begin{equation}
q = {3+\alpha \over 1+\alpha},
\label{eq::alpha.q}
\end{equation}
henceforth, we omit the $q$-Gaussian fit parameter $q$ and explicitly give the fitted values of $\alpha$ and $\beta$ only. For simplicity, we also omit the acronym ``CFD'' and use the asset names only, but one has to realize that the CFD contracts and the assets they refer to are not the same financial entities and that the statistical properties of the former may not necessarily reflect the properties of the latter.


\nointerlineskip

\subsection{Stock Market Indices}

Let us start with the cumulative return distributions of the stock market index CFDs representing six principal indices, NASDAQ100, DJIA, S\&P500, DAX30, FTSE100, and CAC40, and the five time scales, 1 s, 10 s, 1 min, 10 min, and 1 h. Figure~\ref{fig::indices_scales} shows that the return distribution for the three U.S. indices (NASDAQ100, S\&P500, and DJIA) does not show an inverse cubic decay. Dow Jones is the closest, but this may be due to the fact that this index has the least number of aggregated stocks (30 vs. 100 and 500). As the time scale $\Delta t$ increases, we observe a gradual decrease in the thickness of the distribution tails, but this decrease is not so large that a convergence to the normal distribution could firmly be involved. The best power-law fits for $\Delta t = 1$ s are $\alpha \approx 3.9$ (S\&P500), 3.8 (NASDAQ100), and 3.6 (DJIA). For a complete record of the fitted power-law and stretched exponential function parameters, see Table~\ref{tab::indices}. For longer time scales, the tails appear to be significantly thinner only for NASDAQ100, and for $\Delta t = 1$ h, they reach $\alpha \approx 4.6$. For DJIA and S\&P500, we do not observe any convergence to the normal distribution, and therefore, we assume that there is no such convergence for the scales up to 1 h. A strong discrepancy between the inverse cubic and the empirical distribution is also visible in the case of DAX30. For the returns with $\Delta t = 1$ s, we obtain the power law with $\alpha\approx 3.5$, and for the higher scales, we have a trace of $\alpha \to 4$; however, this is by no means a monotonous increase.

\clearpage
\end{paracol}
\nointerlineskip
\begin{specialtable}[H]\setlength{\tabcolsep}{5.05mm}
\widetable
\caption{Estimated tail exponent $\alpha$ and stretched exponent parameter $\beta$ for the aggregated distributions of the CFD returns for  select stock market indices.}
\begin{tabular}{ccccccc}
\toprule
\textbf{Index} & \textbf{Param.} & \boldmath{$\Delta t=1$} \textbf{s} & \boldmath{$\Delta t=10$} \textbf{s} & \boldmath{$\Delta t=1$} \textbf{min} & \boldmath{$\Delta t=10$} \textbf{min} & \boldmath{$\Delta t=1$} \textbf{h} \\ \midrule
DAX30 & $\alpha$ & 3.5 & 3.7 & 3.9 & 3.7 &  2.7 \\
 & $\beta$ & 0.37 & 0.64 & 0.63 & 0.45 & 0.38 \\
CAC40 & $\alpha$ & 3.6 & 3.8 & 3.7 & 3.6 & 4.8 \\
 & $\beta$ & 0.38 & 0.62 & 0.40 & 0.42 & 0.63 \\
FTSE100 & $\alpha$ & 2.8 & 3.4 & 3.7& 3.5 & 4.6 \\
 & $\beta$ & 0.51 & 0.39 & 0.81 & 0.68 & 0.52 \\
DJIA & $\alpha$ & 3.6 & 3.7 & 3.3 & 3.3 & 3.0 \\
 & $\beta$ & 0.37 & 0.41 & 0.68 & 0.41 & 0.37 \\
S\&P500 & $\alpha$ & 3.9 & 3.9 & 3.6 & 3.5 & 3.0 \\
 & $\beta$ & 0.4 & 0.47 & 0.39 & 0.56 & 0.37 \\
NASDAQ100 & $\alpha$ & 3.8 & 4.0 & 3.8 & 3.6 & 4.5 \\
 & $\beta$ & 0.41 & 0.44 & 0.36 & 0.42 & 0.47 \\ \bottomrule
\end{tabular}
\label{tab::indices}
\end{specialtable}
\begin{paracol}{2}
\switchcolumn

The return distribution for the FTSE100 and CAC40 indices are different. Especially in the case of the latter, we observe an approximate inverse cubic decay $\alpha\approx 3$ for $\Delta t = 1$ s. It is also clearer than in the previous cases that the tails become much thinner with increasing scale, and for $\Delta t = 1$ h, we see $\alpha\to 5$. In the case of FTSE100, we do not deal with a homogeneous distribution but, rather, with two or more different distributions imposed. This is visible especially for the shortest time scale, where $\alpha < 3$. As the scale increases, we see a behavior similar to that of CAC40, although it is even more pronounced due to the thicker tail at 1 s.

These results can be compared to those obtained for the high-frequency data from 1998--1999, which included both DJIA and DAX30~\cite{drozdz2003}. The distributions for the shortest scale analyzed ($\Delta t = 5$ min) displayed tails close to those of the inverse cubic ones (even more in the case of DJIA than DAX30), but a crossover was visible for the scales $\Delta t > 2$ h for DJIA and $\Delta t > 30$ min for DAX30. Due to the limited maximum scale considered in the present study, we cannot conclude what the DJIA distributions for the 2 h scale look like, but it seems that, for the shorter scales, these distributions are slightly thinner than before. With regards to the results for the S\&P500 and DAX30 data from the years 2004--2006~\cite{drozdz2007}, the tail slope decrease was power-law starting from $\alpha\approx 4$ for $\Delta t = 1$ min to $\alpha\approx 6$ for $\Delta t = 1$ h for the American index and from $\alpha\approx 3.5$ to $\alpha\approx 5$ for the German index, respectively. These results differ from what we obtained here for the years 2017--2020. It seems that the tendency of the inverse cubic scaling regime to shift towards shorter time scales has at least stopped. The distribution tails also scale worse now than before. However, one has to notice that the years 2004--2006 were characterized by much lower volatility than the years 2017--2020, with a lack of comparably significant, dramatic events, which {can} have some impact on the results.


\nointerlineskip
\begin{figure}[H]

\includegraphics[width=0.7\textwidth]{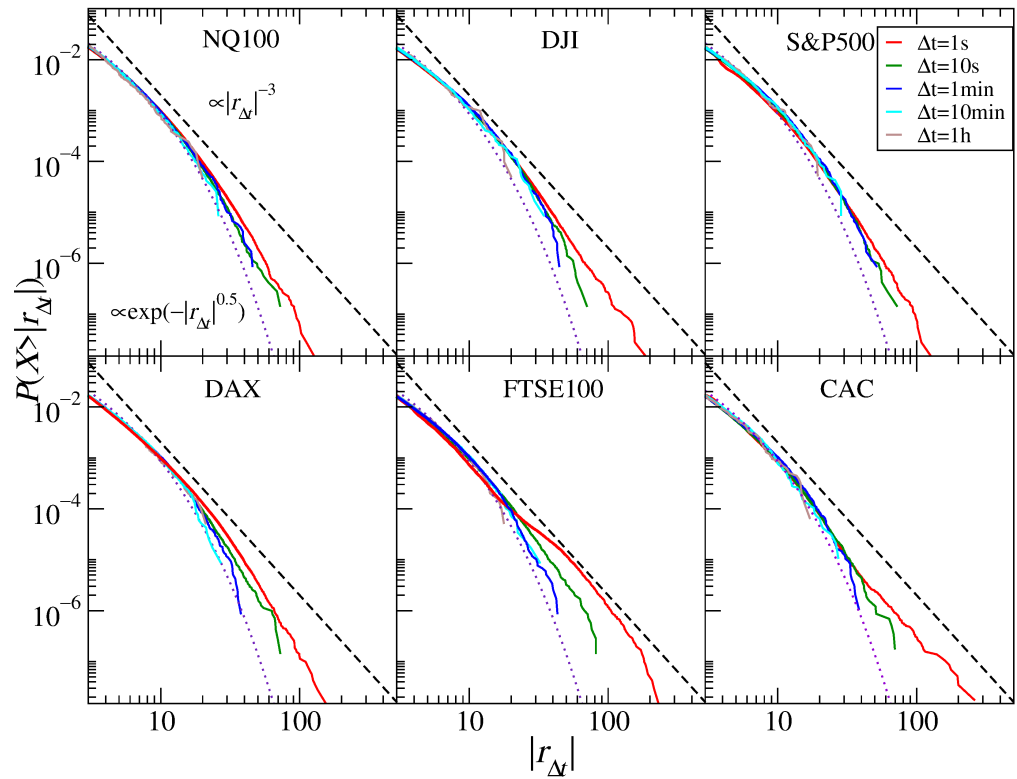}
\caption{Cumulative distribution functions of the CFD returns for stock market indices NQ100 (NASDAQ), DJIA (New York SE), S\&P500 (both New York SE and NASDAQ), DAX30 (Deutsche B\"orse), FTSE100 (London SE), and CAC40 (Euronext). Different sampling intervals (time scales) are shown from 1 s to 1 h. The inverse cubic scaling $\alpha=3$ (dashed line) and the stretched exponential with $\beta=0.5$ (dotted line) are shown in each panel to serve as a guide.}
\label{fig::indices_scales}
\end{figure}

\subsection{Individual Stocks}
\label{sect::stocks}

The return distributions for all individual stocks collected from four mature markets: the U.S., German, British, and French ones are shown in Figure~\ref{fig::largest_stocks_scales}. For the shortest time scale analyzed, three markets display approximate inverse cubic scaling of their tails: $\alpha\approx 3.2$ (U.K. and France) and $\alpha\approx 3.3$ (Germany), while the U.S. market shows a larger exponent: $\alpha\approx 3.6$ (see Table~\ref{tab::stocks}). With increasing $\Delta t$, the distribution tail becomes thinner, and already for $\Delta t=10$ s, the exponent reaches $\approx$3.5 (the European markets) and $\approx $4.0 (the U.S. market). This seems to be the quickest departure from the $\alpha\approx 3$ behavior observed so far for individual stocks. The scaling index increases gradually up to $\Delta t=10$ min, but for 1 h, this picture is altered and only the U.S. stocks show a further increase ($\alpha\approx 5.0$), while the exponent either stops---Germany---or even decreases---the U.K. and France (for these two longest scales, the stretched exponential function fits the empirical distribution better). This makes the situation less clear, but such a non-monotonous behavior was also observed for some scales in~\cite{plerou1999} despite a much larger set of stocks considered there (1000). In that study (the years 1994--1995), $\alpha\approx 5.0$ was observed for the returns sampled every 50--70 trading days. Later studies reported that the scaling regime with $\alpha\approx 3$ already broke at $\Delta t=2$ h for 30 DJIA stocks and at $\Delta t=5$ min for 30 DAX stocks~\cite{drozdz2003} (1998--1999) and then that $\alpha\approx 3$ was valid up to $\Delta t=1$ min and $\alpha\approx 5$ was reached for $\Delta t=2$ h~\cite{drozdz2007} (1000 U.S. stocks, 1998--1999). In fact, even though in Table~\ref{tab::stocks} we do not observe a convincing convergence of the empirical distributions for the European stocks towards the normal distribution, our results show that contemporary stocks experienced an accelerated time flow  compared with the past. Of course, since we analyzed the CFD contracts instead of the stock share spot quotations as in~\cite{plerou1999,drozdz2003,drozdz2007}, we have to be careful in drawing decisive conclusions from the comparison between these two asset types.


\nointerlineskip
\begin{figure}[H]
\widetable
\includegraphics[width=0.7\textwidth]{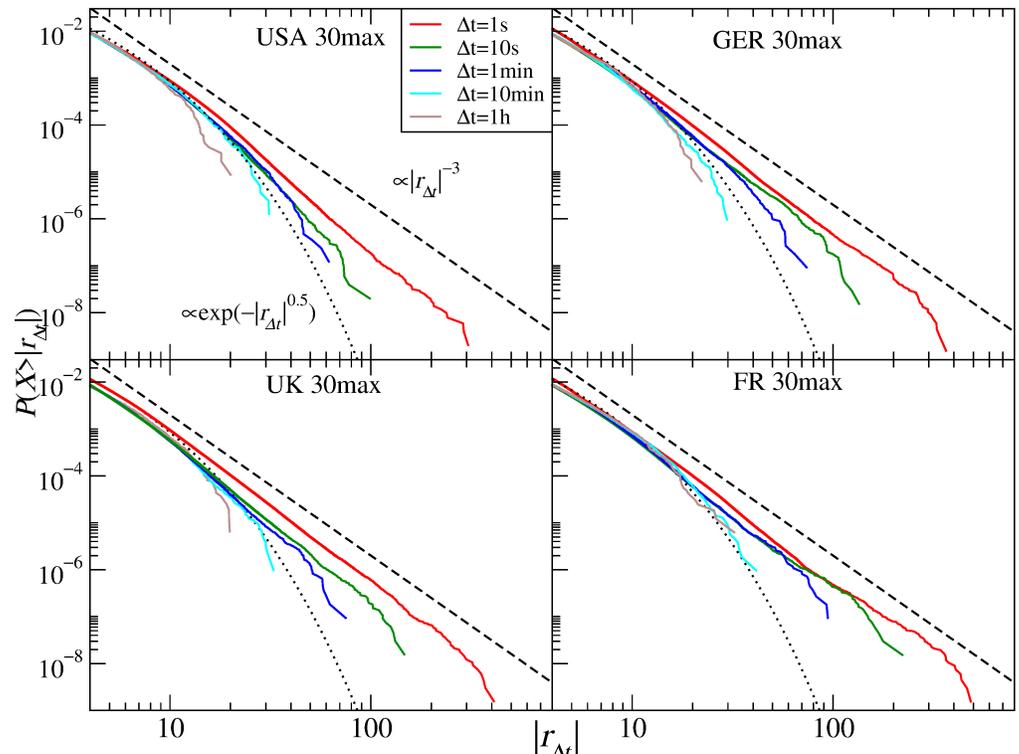}
\caption{Cumulative distribution functions of the CFD returns for the stock shares representing different markets: the U.S. market (USA), the German market (GER), the U.K. market (UK), and the French market (FR). In each case, the aggregated distributions for 30 stocks with the largest capitalization are shown. Different time scales are shown from 1 s to 1 h. The inverse cubic scaling $\alpha=3$ (dashed line) and the stretched exponential with $\beta=0.5$ (dotted line) are shown in each panel to serve as a guide.}
\label{fig::largest_stocks_scales}
\end{figure}

The faster convergence of aggregated returns nowadays, compared with a more or less distant past, is among others a consequence of a decreasing autocorrelation time~\cite{plerou1999,drozdz2003,drozdz2007}. On the other hand, somehow, an opposite process is the increase in the cross-correlation magnitude among different stocks, which leads to stronger violation of the CLT assumption about random variable independence and thickening of distribution tails for the stock market indices, which can cause a later crossover to the CLT regime. For the shortest scales available for analysis of the order of an inter-trade interval, the cross-correlations are relatively weak due to strong noise and a longer time needed for information to spread over the market. However, by increasing $\Delta t$, we also increase the cross-correlation magnitude, which can eventually reach a saturation level with a magnitude dependent on the stocks considered (the same industrial sector vs. different sectors, whether the stocks are included in the same index, etc.)~\cite{epps1979,kwapien2004,drozdz2010,kwapien2012}. If we review the available results on this problem, we can see that, in 1971, the saturation of the cross-correlation coefficient for the stocks of the largest capitalization was reached at \mbox{$\Delta t\approx 1$ day~\cite{epps1979}}, while  it was 1/2 h for 1998--1999 for the largest companies and a few hours for the medium-sized companies~\cite{kwapien2004}. In order to learn how much time is needed for the cross-correlation magnitude to saturate nowadays, we calculated the Pearson cross-correlation coefficients $C_{ij}(\Delta t)$ ($i,j=1,...,30$) for all pairs of stocks within each of the markets studied here. Figure~\ref{fig::epps_scales} shows the results for the mean coefficient $\langle C_{ij}\rangle$ together with a mean length of the zero-return sequences in the analyzed time series and the largest eigenvalue of the $30 \times 30$ correlation matrix ${\bf C}(\Delta t)$ in which the elements are $C_{ij}$ for a given $\Delta t$ and a given market. We also added two sets of U.S. stocks that represent medium-sized and small capitalization stocks. For all sets of stocks, a trace of saturation is observed already at the time scales of a few minutes, which is much less than the numbers presented above that from earlier works. This validates our statement that the market time ``felt'' by the assets accelerates. There is also a clear dependence of the mean cross-correlation coefficient on stock capitalization: the larger the capitalization, the stronger the correlation \mbox{Figure~\ref{fig::epps_scales}.}

\end{paracol}
\nointerlineskip
\begin{specialtable}[H]\setlength{\tabcolsep}{5mm}
\widetable
\caption{Estimated tail exponent $\alpha$ for the aggregated distributions of the CFD returns for 30 U.S. stocks with the largest, medium, and small capitalizations and 30 stocks representing {selected} European markets.}
\begin{tabular}{ccccccc}
\toprule
\textbf{Market }& \textbf{Param.} & \boldmath{$\Delta t=1$} \textbf{s} & \boldmath{$\Delta t=10$} \textbf{s} & \boldmath{$\Delta t=1$} \textbf{min} & \boldmath{$\Delta t=10$} \textbf{min} & \boldmath{$\Delta t=10$} \textbf{h} \\ \midrule
U.S. large & $\alpha$ & 3.7 & 4.0 & 3.9 & 4.0 & 5.0 \\
 & $\beta$ & & & & 0.47 & 0.54 \\
U.S. medium & $\alpha$ & 3.7 & 4.1 & 3.8 & 4.0 & 4.0 \\
 & $\beta$ & & & & 0.41 & 0.48 \\
U.S. small & $\alpha$ & 3.5 & 3.7 & 3.8 & 3.9 & 4.5 \\
 & $\beta$ & & & & 0.46 & 0.56 \\
Germany & $\alpha$ & 3.3 & 3.6 & 3.8 & 4.2 & 4.2 \\
 & $\beta$ & & & & 0.49 & 0.56 \\
U.K. & $\alpha$ & 3.2 & 3.5 & 3.9 & 4.2 & 3.6 \\
 & $\beta$ & & & & 0.51 & 0.48 \\
France & $\alpha$ & 3.2 & 3.4 & 3.5 & 3.7 & 3.5 \\
 & $\beta$ & & & & 0.50 & 0.44 \\ \bottomrule
\end{tabular}
\label{tab::stocks}
\end{specialtable}
\begin{paracol}{2}
\switchcolumn

It is well-known that the cross-correlations are not stationary and that they strongly fluctuate across time~\cite{drozdz2001b,drozdz2002,kwapien2012}. Figure~\ref{fig::pearson_evolution} displays the evolution of $\langle C_{ij}(t) \rangle$ calculated in 30-day windows over the years 2018--2020. Two time scales are considered: 1 s and 1 h. $\langle C_{ij}(t) \rangle$ fluctuates with a larger amplitude for $\Delta t=1$ h than for $\Delta t=1$ s. One of the periods associated with the largest values of $\langle C_{ij}(t)\rangle$ is 9--27 March 2020 (the COVID-19 pandemic outburst in the U.S.), when the markets underwent strong turbulence~\cite{watorek2021}. As the cross-correlations were particularly strong during that period, we suspect that it could contribute substantially to the tail shape of the stock return distributions.

To verify this hypothesis, we removed this period from the time series and constructed artificial stock indices by aggregating the returns for all stocks belonging to the same set. Figure~\ref{fig::art_indices_scales} shows both the complete distributions and the resultant no-COVID ones. After removing the COVID-19 outburst period, the distribution tails became substantially thinner, which is particularly evident for $\Delta t=1$ h (see Table~\ref{tab::nocovid}). This supports our hypothesis that strong cross-correlations among the stocks can prevent stock indices from showing CLT convergence for short time scales. In this case, the stretched exponential function fits the empirical distribution better than the power-law function (Table~\ref{tab::nocovid}). The numbers in this table illustrate how the stock-stock correlation strength can influence the stock index returns. While the stretching parameter $\beta$ is comparable for each group of the U.S. stocks at $\Delta t=1$ s, it becomes significantly different at $\Delta t=1$ h, where the medium and small companies have thinner tails than the large companies. This is because the former are less cross-correlated than the latter and the distributions can more easily converge towards a Gaussian in this case, even though the medium and small companies should experience a slower time flow than the large ones, which acts towards tail thickening. From this example, we can see that both effects compete against each other and that the actual tail behavior depends on the interplay of both factors.


\nointerlineskip
\begin{figure}[H]
\includegraphics[width=0.7\textwidth]{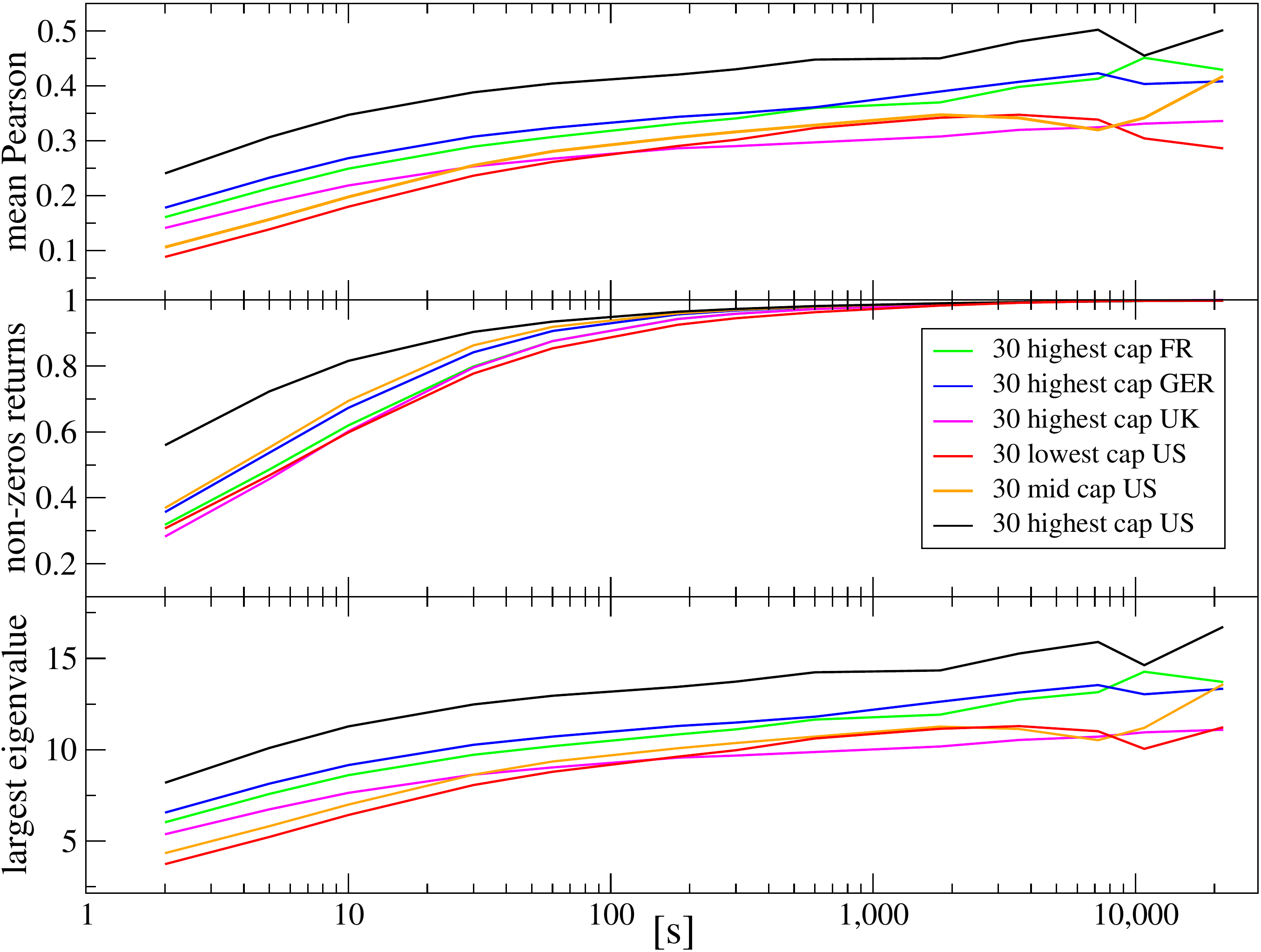}
\caption{(\textbf{Top}) The mean Pearson cross-correlation coefficient $\langle C_{ij} \rangle(\Delta t)$ for the CFD returns as a function of time scale $\Delta t$ for 30 companies, with the largest capitalization representing four stock markets, French (FR), German (GER), British (UK), and American (US), and for 30 companies with medium and small capitalization from the American market. Averaging was carried out over all pairs $i,j$ with $i>j$ and $i,j=1,...,30$. (\textbf{Middle}) The same was performed as above, but here, the zero returns were filtered out before calculating the correlation coefficients. (\textbf{Bottom}) The largest eigenvalue of the correlation matrix ${\bf C}(s)$ constructed from the Pearson cross-correlation coefficients $C_{ij}(s)$ for the same sets of stock share CFDs.}
\label{fig::epps_scales}
\end{figure}

`

\nointerlineskip
\begin{figure}[H]
\widefigure
\includegraphics[width=0.7\textwidth]{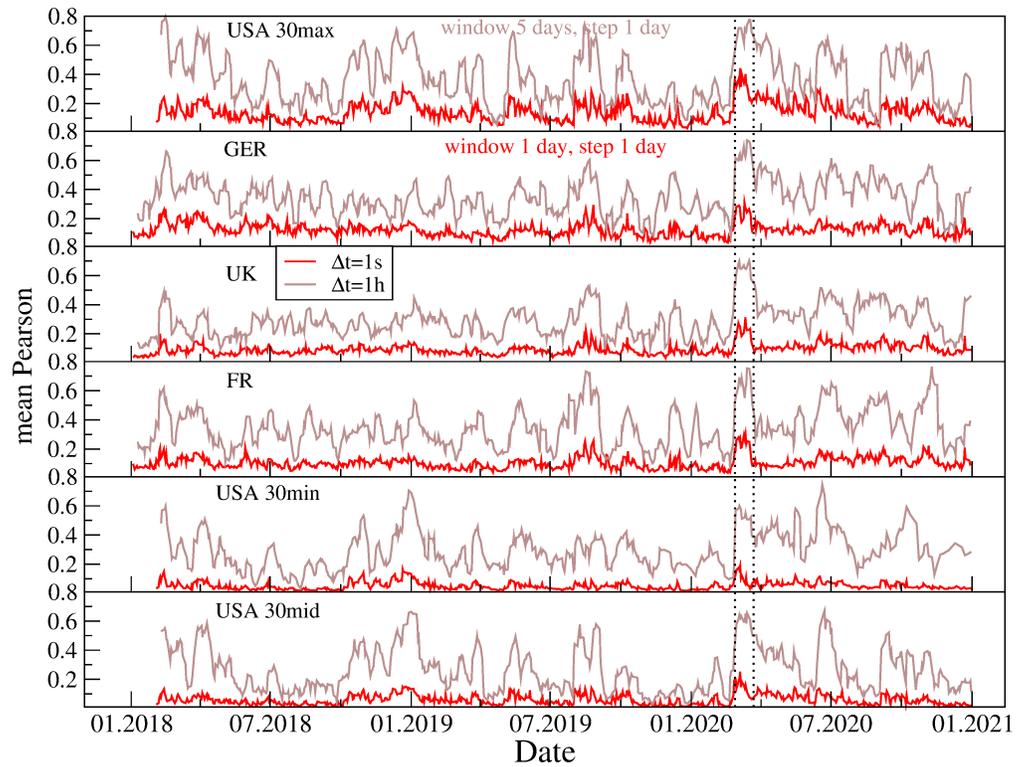}
\caption{Evolution of the mean Pearson cross-correlation coefficient $\langle C_{ij} \rangle(\Delta t)(t)$ for the CFD returns of 30 companies, with the largest capitalization representing four stock markets, French (FR), German (GER), British (UK), and American (US), and for 30 companies with medium and small capitalization from the American market. The coefficient was calculated in a moving window of length of 30 days, and  averaging was carried out over all pairs $i,j$ with $i>j$ and $i,j=1,...,30$. Two time scales with $\Delta t=1$ s and $\Delta t = 1$ h are shown in each case.}
\label{fig::pearson_evolution}
\end{figure}

\end{paracol}
\nointerlineskip
\begin{specialtable}[H]\setlength{\tabcolsep}{5mm}
\widetable
\caption{Estimated tail exponent $\alpha$ and stretched exponent parameter $\beta$ for the aggregated distributions of the CFDs returns for 30 artificial stock indices representing different markets and different stock capitalization groups from the U.S. market. The results for the complete data and the data without the COVID-19 outburst in March 2020 (denoted nC) are shown \mbox{for comparison.}}
\begin{tabular}{ccccccc}
\toprule
\textbf{Market} &\textbf{ \# Stocks} &\textbf{ Param.} & \boldmath{$\Delta t=1$} \textbf{s} & \boldmath{$\Delta t=1$} \textbf{s (nC)} & \boldmath{$\Delta t=1$} \textbf{h} & \boldmath{$\Delta t=1$} \textbf{h (nC)} \\ \midrule
U.S. large & 30 & $\alpha$ & 5.5 & 5.1 & 3.0 & 5.7 \\
 & & $\beta$ & 0.51 & 0.49 & 0.49 & 0.56 \\
U.S. mid & 30 & $\alpha$ & 4.7 & 4.4 & 2.7 & 6.6 \\
 & & $\beta$ & 0.43 & 0.41 & 0.48 & 0.72 \\
U.S. small & 30 & $\alpha$ & 4.4 & 4.4 & 5.2 & 5.5 \\
 & & $\beta$ & 0.43 & 0.41& 0.74 & 0.76 \\
Germany & 30 & $\alpha$ & 3.7 & 3.8 & 2.3 & 3.9 \\
 & & $\beta$ & 0.45 & 0.43 & 0.32 & 0.50 \\
U.K. & 30 & $\alpha$ & 4.0 & 4.6 & 2.7 & 4.7 \\
 & & $\beta$ & 0.41 & 0.42 & 0.46 & 0.66 \\
France & 30 & $\alpha$ & 3.9 & 4.0 & 2.7 & 6.2 \\
 & & $\beta$ & 0.41 & 0.42 & 0.41 & 0.74 \\ \bottomrule
\end{tabular}
\label{tab::nocovid}
\end{specialtable}
\begin{paracol}{2}
\switchcolumn


\nointerlineskip
\begin{figure}[H]

\includegraphics[width=0.7\textwidth]{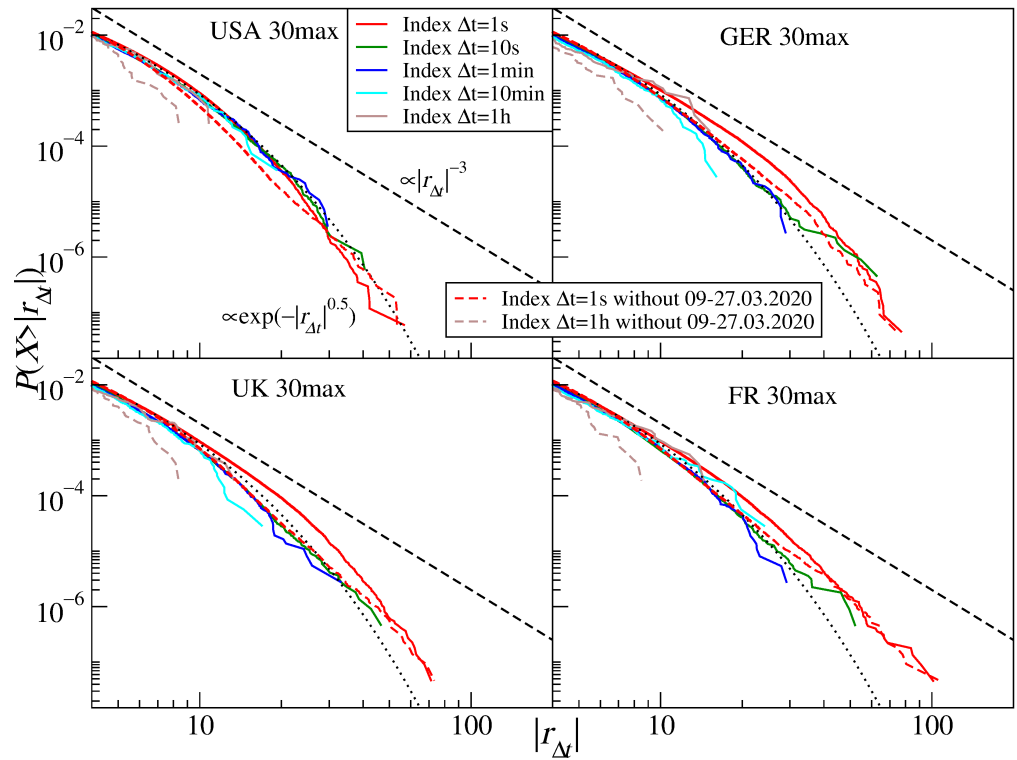}
\caption{Cumulative distribution functions of the returns of an artificial index constructed as a sum of the stock share CFD quotes $I(t) = \sum_{i} P_i(t)$ for the 30 largest companies  representing four stock markets, the U.S. market (USA), the German market (GER), the U.K. market (UK), and the French market (FR). Two time scales with $\Delta t=1$ s and $\Delta t = 1$ h are shown and denoted by solid lines. In addition, the analogous distributions constructed from the CFD return time series after removing the COVID-19 outburst period corresponding to the strongest cross-correlations among the stock shares (9--27 March 2020) are denoted by dashed lines. The inverse cubic scaling $\alpha=3$ (dashed line) and the stretched exponential with $\beta=0.5$ (dotted line) are shown in each panel to serve as a guide to \mbox{the eye.}}
\label{fig::art_indices_scales}
\end{figure}

\subsection{Currencies}

Unlike the stock indices, the return distributions for the exchange rates show the presence of increasingly thinner tails if the time scales increase, and hence, a faster convergence towards the normal distribution (Figure~\ref{fig::forex_scales}). If fitted by a power function, the differences between the individual exchange rates are smaller than those of the indices and generally exhibit an inverse cubic decay for smaller $\Delta t$s: for $\Delta t = 1 $ s, they vary from $\alpha\approx 3.0$ (USD/JPY, GBP/USD) to $\alpha\approx 3.4$ (EUR/USD) and, for $\Delta t = 1$ h, from $\alpha\approx 4.2$ (EUR/JPY) to $\alpha\approx 5.5$ (GBP/CHF) with the exception of GBP/JPY ($\alpha\approx 2.8$). The numbers are collected in Table~\ref{tab::currencies}. The scaling exponent $\alpha\approx 3$ was observed in many studies of the Forex data, including References~\cite{guillaume1997,drozdz2010,drozdz2019,gebarowski2019,watorek2021}. If the stretched exponential function is used, the best-fitted parameter $\beta$ reads for $\Delta t = 1$ s from $\beta = 0.37$ (EUR/JPY) to $\beta = 0.48$ (GBP/JPY, USD/CHF) and, for $\Delta t = 1$ h, from $\beta = 0.49$ (EUR/USD, USD/JPY) to $\beta = 0.87$ (GBP/USD). The mean values of the scaling exponent for the analyzed time scales are $\bar{\alpha} = 3.2$ (1 s), $\bar{\alpha} = 3.1$ (10 s), $\bar{\alpha} = 3.2$ (1 min), $\bar{\alpha} = 3.7$ (10 min), and $\bar{\alpha} = 5.0$ (1 h). The inverse cubic scaling can therefore now be identified for scales shorter than 10 min. These scale are longer than those in the years 2004--2008 for the 1-min scale $\bar{\alpha} = 3.9$~\cite{drozdz2010}. (It has to be noted, however, that those earlier results were obtained by fitting the $q$-Gaussian functions instead of the power-law functions, which may make it difficult to compare the results properly even though the relation {given by} Equation (\ref{eq::alpha.q}) holds.) We thus observe a slower convergence to the normal distribution now than before for $\Delta t = 1$ h: $\bar{\alpha} = 5.9$ (2004--2008) vs. $\bar{\alpha} = 4.8$ (2017--2020). However, if the obtained results are compared with those from the study~\cite{guillaume1997} for the years 1987--1993, the acceleration becomes visible: $\bar{\alpha} = 3.9$ (1987--1993) vs. $\bar{\alpha} = 4.8$ (2017--2020). In that case, the inverse cubic scaling was still visible for the 30-min scale, which is much longer than both in 2004--2008 and 2017--2020. This can be interpreted as the acceleration of the market time resulting in a faster convergence to the normal distribution in the 2000s, but later, this phenomenon of the effective time scale shortening disappeared.


\nointerlineskip
\begin{figure}[H]

\includegraphics[width=0.7\textwidth]{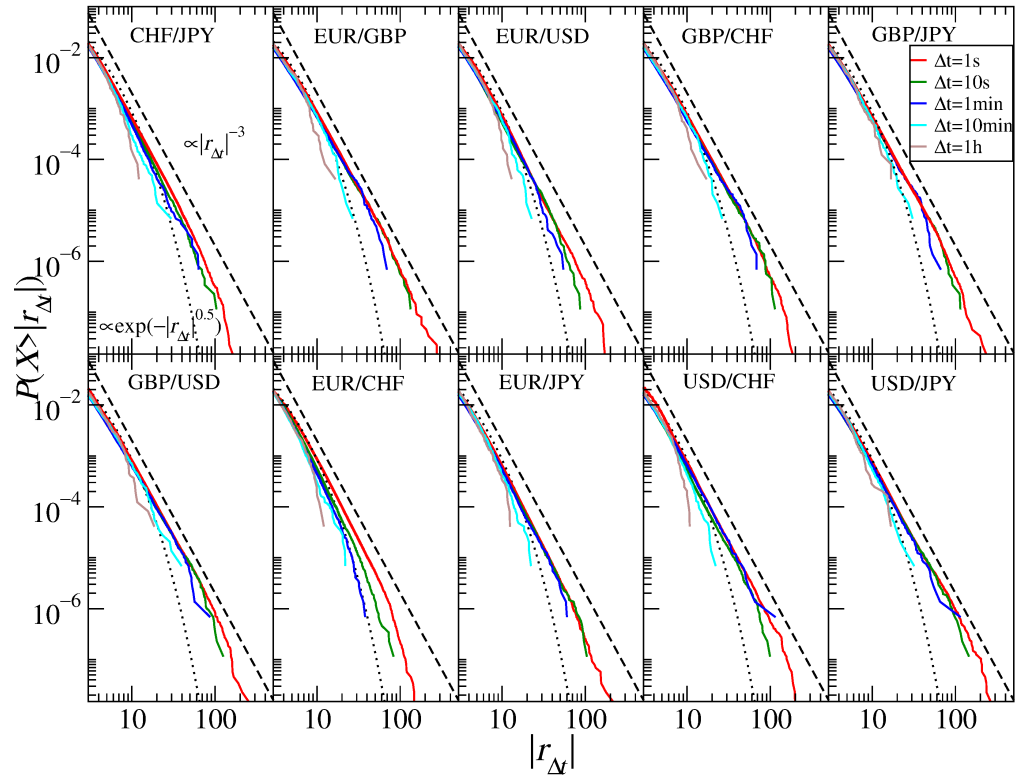}
\caption{Cumulative distribution functions of the major currency exchange rate returns: Swiss franc (CHF), euro (EUR), British pound (GBP), Japanese yen (JPY), and the U.S. dollar (USD). Different time scales are shown from 1 s to 1 h. The inverse cubic scaling $\alpha=3$ (dashed line) and the stretched exponential with $\beta=0.5$ (dotted line) are shown in each panel to serve as a guide.}
\label{fig::forex_scales}
\end{figure}


\clearpage
\end{paracol}
\nointerlineskip
\begin{specialtable}[H]\setlength{\tabcolsep}{4.9mm}
\widetable
\caption{Estimated tail exponent $\alpha$ and stretched exponent parameter $\beta$ for the aggregated return distributions for the selected currency exchange rates and the cryptocurrency prices: BTC/USD and ETH/USD.}
\begin{tabular}{ccccccc}
\toprule
\textbf{Exchange Rate} & \textbf{Param.} & \boldmath{$\Delta t=1$} \textbf{s} & \boldmath{$\Delta t=10$} \textbf{s} & \boldmath{$\Delta t=1$} \textbf{min} & \boldmath{$\Delta t=10$} \textbf{min} & \boldmath{$\Delta t=1$} \textbf{h} \\ \midrule
CHF/JPY & $\alpha$ & 3.2 & 3.3 & 3.4 & 3.6 &  5.2 \\
 & $\beta$ & 0.48 & 0.37 & 0.51 & 0.41 & 0.61 \\
EUR/CHF & $\alpha$ & 3.3 & 3.6 & 4.0 & 3.9 & 5.2 \\
 & $\beta$ & 0.40 & 0.39 & 0.41 & 0.40 & 0.78 \\
EUR/GBP & $\alpha$ & 3.1 & 2.9 & 2.9 & 3.6 & 5.1 \\
 & $\beta$ & 0.42 & 0.34 & 0.33 & 0.48 & 0.75 \\
EUR/JPY & $\alpha$ & 3.3 & 3.1 & 3.1 & 4.0 & 4.2 \\
 & $\beta$ & 0.37 & 0.29 & 0.39 & 0.64 & 0.58 \\
EUR/USD & $\alpha$ & 3.4 & 3.2 & 3.4 & 4.4 & 5.2 \\
 & $\beta$ & 0.39 & 0.35 & 0.37 & 0.55 & 0.65 \\
GBP/CHF & $\alpha$ & 3.1 & 2.9 & 2.9 & 3.5 & 5.5 \\
 & $\beta$ & 0.40 & 0.36 & 0.40 & 0.42 & 0.50 \\
GBP/JPY & $\alpha$ & 3.1 & 2.9 & 3.0 & 3.4 & 2.8 \\
 & $\beta$ & 0.48 & 0.34 & 0.36 & 0.54 & 0.37 \\
GBP/USD & $\alpha$ & 3.0 & 2.9 & 2.9 & 3.3 & 5.1 \\
 & $\beta$ & 0.40 & 0.33 & 0.34 & 0.39 & 0.87 \\
USD/CHF & $\alpha$ & 3.2 & 3.1 & 3.3 & 4.1 & 5.2 \\
 & $\beta$ & 0.45 & 0.52 & 0.38 & 0.48 & 0.62 \\
USD/JPY & $\alpha$ & 3.0 & 2.9 & 3.0 & 3.5 & 5.2 \\
 & $\beta$ & 0.40 & 0.34 & 0.36 & 0.56 & 0.54 \\ 
BTC/USD & $\alpha$ & 2.9 & 3.1 & 3.2 & 3.2 & 3.7 \\
 & $\beta$ & & & & & \\
ETH/USD & $\alpha$ & 2.8 & 3.1 & 3.2 & 3.3 & 4.3 \\
 & $\beta$ & & & & & 0.50 \\ \bottomrule
\end{tabular}
\label{tab::currencies}
\end{specialtable}
\begin{paracol}{2}
\switchcolumn

\subsection{Cryptocurrencies}

The cryptocurrency market is strongly related to Forex and its significance has risen steadily since its beginning~\cite{drozdz2018,watorek2021}. The most important assets traded on this market in terms of their capitalization and volume are bitcoin (BTC) and ethereum (ETH). Their return distributions are shown in Figure~\ref{fig::crypto_scales}. For $\Delta t$ up to 10 min, the power-law function approximates the data well, with the tail exponent displaying the same inverse cubic scaling for BTC and ETH: $\alpha\approx 2.8$ (1 s), $\alpha\approx 3.1$ (10 s), $\alpha\approx 3.2$ (1 min), $\alpha\approx 3.3$ (10 min). In contrast, for $\Delta t=1$ h, the crossover is observed and the tail exponent rises to  $\alpha\approx 3.7$ for BTC and to 4.2 for ETH. (The stretched exponential function does not fit the data in the tail region, except on the 1 h scale for ETH) A good agreement between the empirical distributions of the cryptocurrency price returns expressed in major regular currencies and the inverse cubic scaling paradigm has already been reported \cite{drozdz2018,drozdz2019,watorek2021,takaishi2021}, and it was interpreted as a sign of maturation (it used to be even more heavy-tailed with $\alpha\approx 2.2$ before 2014~\cite{begusic2018,drozdz2018}). A crossover for the same scale $\Delta t=1$ h has also been reported~\cite{drozdz2019}.

A more time-resolution-oriented analysis~\cite{drozdz2020a} showed that the BTC dynamics can actually be compounded with alternating phases of fluctuations with different statistical properties. For example, during the COVID-19 outburst in March 2020, the BTC returns were characterized by $\alpha\approx 1.8$, which corresponds to the L\'evy stability, but typically, the scaling index resided between 2.0 and 3.5 during the years 2019--2020~\cite{drozdz2020a}. In contrast, the L\'evy-stable distribution of daily returns with $\alpha\approx 1.3$ was reported to fit the BTC empirical data (the years 2011--2017) in~\cite{muvunza2020}, but there, the model was fitted to the whole distributions, not only the tails, and the analyzed period also covered the early years of the cryptocurrency market when it was immature. These may be a source of the discrepancy mentioned above.


\nointerlineskip
\begin{figure}[H]

\includegraphics[width=0.7\textwidth]{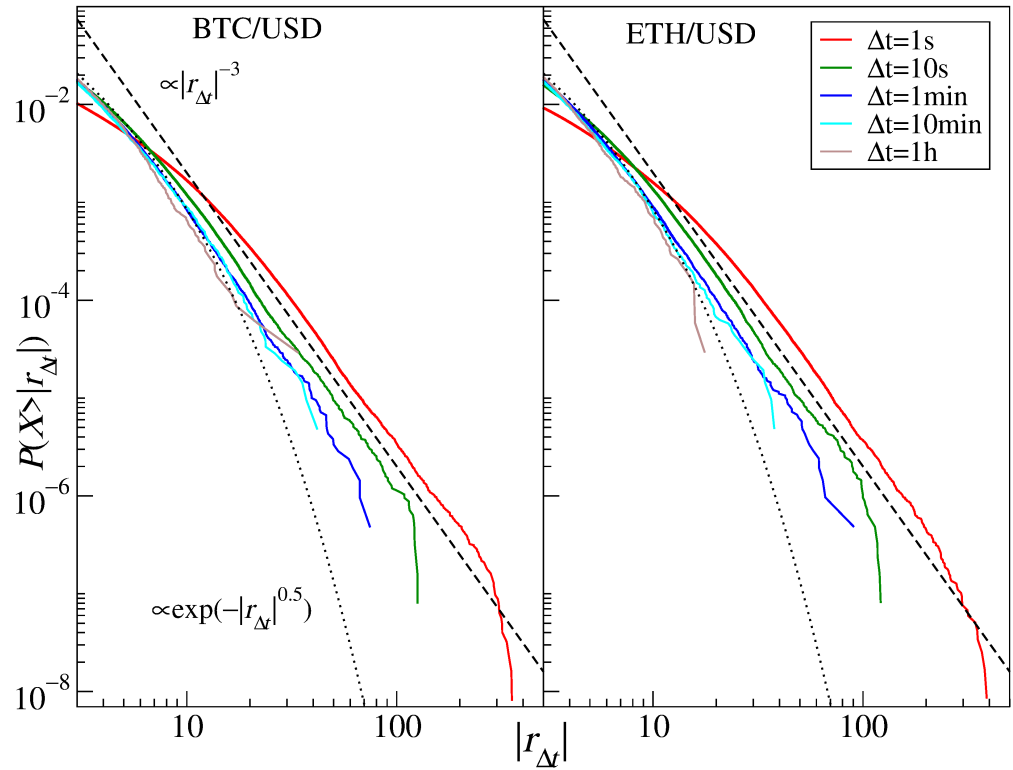}
\caption{Cumulative distribution functions of the bitcoin-dollar exchange rate returns (BTC/USD) and the ethereum-dollar exchange rate returns (ETH/USD). Different time scales are shown from 1 s to 1 h. The inverse cubic scaling $\alpha=3$ (dashed line) and the stretched exponential with $\beta=0.5$ (dotted line) are shown in each panel to serve as a guide.}
\label{fig::crypto_scales}
\end{figure}

\subsection{Commodities}

Figure~\ref{fig::commodities_scales} shows the return distributions for the commodity CFDs (see also Table~\ref{tab::commodities}). Out of the commodities considered in our study, gold (XAU) has the strongest decay with increasing $\Delta t$: $\alpha\approx 2.6$ (1 s), $\alpha\approx 3.1$ (10 s), $\alpha\approx 3.6$ (1 min), $\alpha\approx 3.6$ (10 min), and $\alpha\approx 4.2$ (1 h). This has to be compared to $\alpha\approx 2.5$ for daily returns covering the years 1969-1999~\cite{matia2002} and 5-min returns covering the years 2012--2018~\cite{watorek2019}. It is instructive to address why the tail for $\Delta t=1$ s is so thick that $\alpha$ falls significantly below 3. We therefore identified a period of the largest gold price fluctuations, which occurred at the COVID-19 outburst in the U.S. during March and April, and removed it from the time series. The resulting distributions are shown in Figure~\ref{fig::gold_oil_no-covid} (left panel) for $\Delta t=1$ s, 1 min, and 1 h (dashed lines) together with the complete ones (solid lines). It is evident that, for $\Delta t=1$ s, the thickest part of the tail becomes thinner $\alpha\approx 3.2$, while no significant alternation is observed for $\Delta t=1$ min ($\alpha\approx 3.6$) and 1 h ($\alpha\approx 4.2$). The distribution tails roughly agree in this case with the inverse cubic scaling on time scales that become increasingly short with time.

Compared to gold, silver (XAG) shows stronger invariance under the time scale change: $\alpha$ increases from 2.8 (1 s) to 3.0 (1 h), while it was 2.5 (positive tail) and 2.8~(negative tail) for the daily returns over the years 1969--1999~\cite{matia2002}. High-grade copper return distributions (HG) reveal the most wandering behavior: its scaling exponent goes from $\alpha\approx 3.7$ (1 s) through $\alpha\approx 3.6$ (10 s), $\alpha\approx 3.1$ (1 min), and $\alpha\approx 3.9$ (10 min) to $\alpha\approx 2.7$ (1 h) compared to $\alpha\approx 2.6$ (negative tail; daily data) and $\alpha\approx 2.8$ (positive tail) for the years 1971--1999~\cite{matia2002}.

\clearpage
\end{paracol}
\nointerlineskip
\begin{specialtable}[H]\setlength{\tabcolsep}{5.1mm}
\widetable
\caption{Estimated tail exponent $\alpha$ and stretched exponent parameter $\beta$ for the aggregated distributions of the CFDs returns for the selected commodities: high-grade copper (HG), crude oil (CL), silver (XAG), and gold (XAU).}
\begin{tabular}{ccccccc}
\toprule
\textbf{Commodity} & \textbf{Param. }& \boldmath{$\Delta t=1$} \textbf{s} & \boldmath{$\Delta t=10$} \textbf{s} & \boldmath{$\Delta t=1$} \textbf{min} & \boldmath{$\Delta t=10$} \textbf{min} & \boldmath{$\Delta t=1$} \textbf{h} \\ \midrule
HG & $\alpha$ & 3.8 & 3.6 & 3.1 & 3.9 &  2.7 \\
 & $\beta$ & 0.42 & 0.49 & 0.36 & 0.45 & 0.25 \\
CL & $\alpha$ & 2.6 & 2.3 & 2.3 & 2.0 & 2.5 \\
 & $\beta$ & 0.48 & 0.29 & 0.28 & 0.41 & 0.43 \\
XAG & $\alpha$ & 2.8 & 2.9 & 2.9 & 3.0 & 3.0 \\
 & $\beta$ & 0.47 & 0.36 & 0.37 & 0.38 & 0.43 \\
XAU & $\alpha$ & 2.6 & 3.1 & 3.6 & 3.6 & 4.2 \\
 & $\beta$ & 0.49 & 0.62 & 0.42 & 0.51 & 0.85 \\ \bottomrule
\end{tabular}
\label{tab::commodities}
\end{specialtable}
\begin{paracol}{2}
\switchcolumn

\begin{figure}[H]
\includegraphics[width=0.7\textwidth]{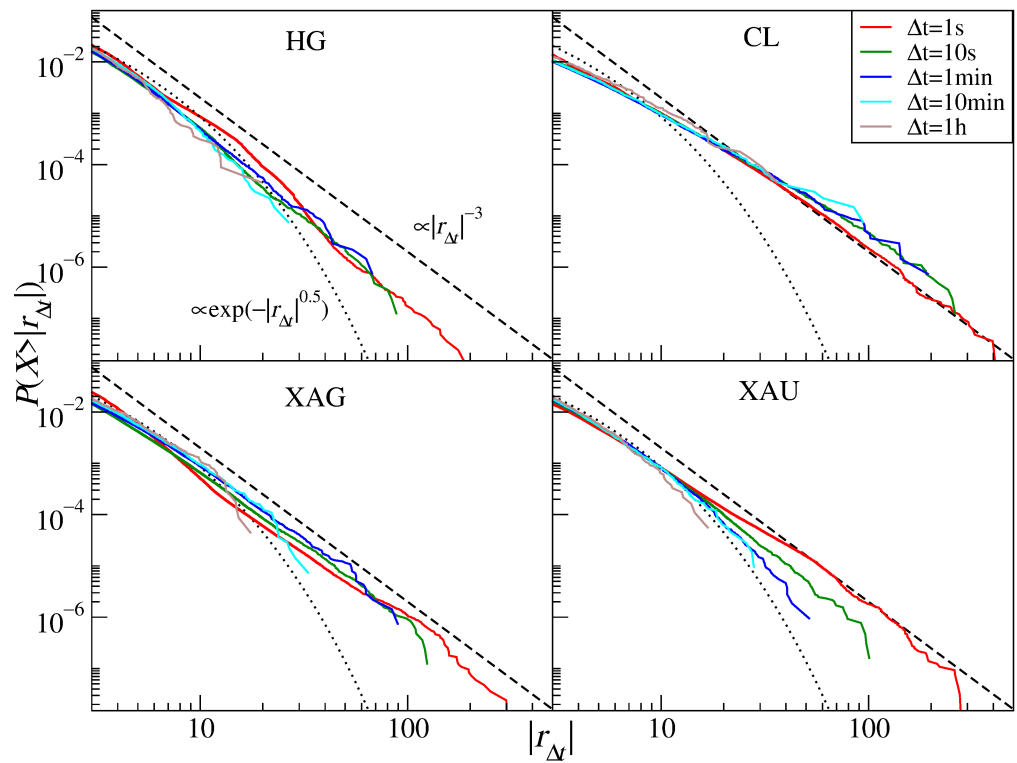}
\caption{Cumulative distribution functions of the CFD returns for commodities: high-grade copper (HG), the U.S. crude oil (CL), silver (XAG), and gold (XAU). Different time scales are shown from 1 s to 1 h. The inverse cubic scaling $\alpha=3$ (dashed line) and the stretched exponential with $\beta=0.5$ (dotted line) are shown in each panel to serve as a guide.}
\label{fig::commodities_scales}
\end{figure}

Crude oil return distributions (CL) have the thickest tails with the scaling exponent $2.0 \le \alpha \le 2.6$ and with no signature of the CLT convergence. In parallel to what was observed for gold, we removed the COVID-19 outburst period from the time series and calculated the distributions again---see Figure~\ref{fig::gold_oil_no-covid} (right panel). Such incomplete signals are characterized by $\alpha=3.0$ for $\Delta t=1$ s and $\alpha\approx 2.7$ for $\Delta t=1 $ h. These numbers have to be compared with $\alpha\approx 2.9$ (negative tail) and $\alpha\approx 3.1$ (positive tail) reported for the WTI oil daily returns covering the years 1988--1998, with $\alpha\approx 2.0$ (negative tail) and $\alpha\approx 2.8$ (positive tail) reported for the crude oil daily returns covering the years 1983--1999~\cite{matia2002}, and with $\alpha\approx 3.0$ (negative tail) and $\alpha\approx 3.1$ (positive tail) for the WTI oil 5-min returns covering the years 2012--2018~\cite{watorek2019}. As those values do not differ much from each other, there is no evidence that the crude oil returns change their global dynamics over time. However, during the market turbulence that happened during the COVID-19 pandemic, the dynamics did change considerably, which was manifested by thickening the distribution tail for all considered time scales.  It is noteworthy that the crude oil  was the asset that was affected the strongest by the pandemic: in April 2020, the price of the May series of WTI oil futures  even dropped below 0.

\begin{figure}[H]
\includegraphics[width=0.7\textwidth]{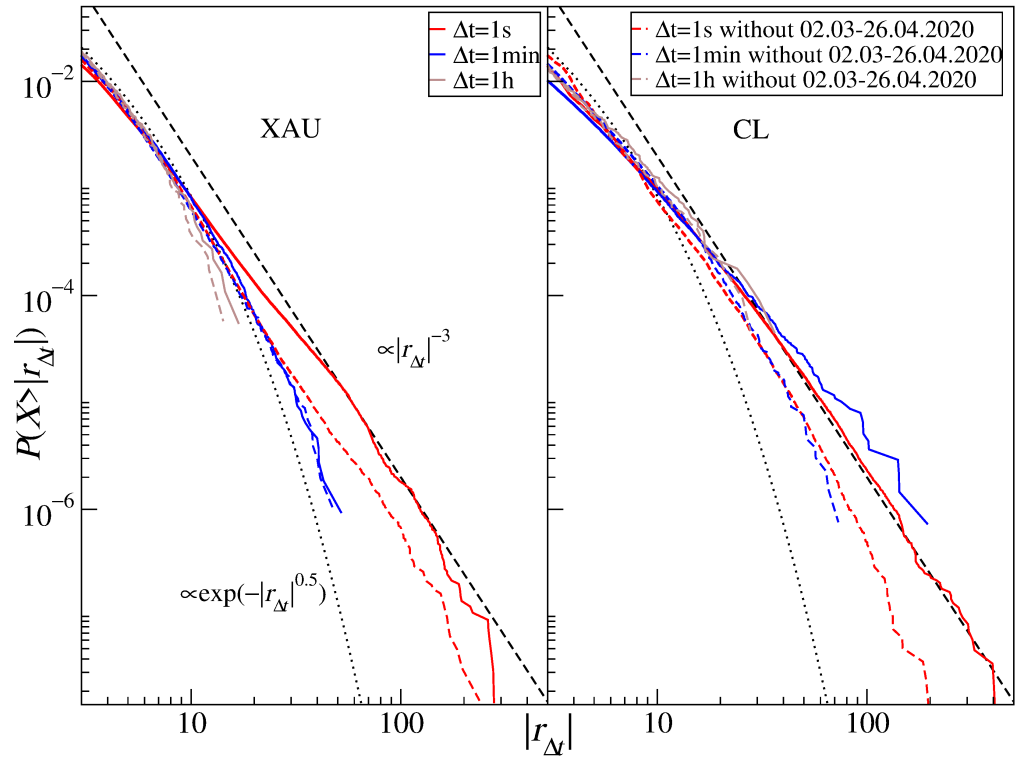}
\caption{Cumulative distribution functions of the CFD returns for gold (XAU) and the U.S. crude oil (CL) after removing the COVID-19 outburst period (March--April 2020). Different time scales are shown from 1 s to 1 h. The inverse cubic scaling $\alpha=3$ (dashed line) and the stretched exponential with $\beta=0.5$ (dotted line) are shown in each panel to serve as a guide.}
\label{fig::gold_oil_no-covid}
\end{figure}

\section{Summary}\label{sec4}

In this study, we analyzed high-frequency quotations of the CFD contracts associated with the stock market indices, the stocks themselves, and the selected commodities as well as with the most frequently traded currency exchange rates and the cryptocurrency prices. All of the data sets covered the years 2017--2020 except for the stock share CFDs, which covered the years 2018--2020. We analyzed the returns at a few different time scales from 1 s to 1 h and constructed the return distributions in order to investigate their tails. Our principal objective was to compare the tail behavior of the distributions derived from contemporary data with the behavior of the distribution tails in the past for the same assets. We applied the power-law function and the stretched exponential one to model the empirical distributions. A hypothesis that we planned to verify was the one formulated in~\cite{drozdz2003,drozdz2007,kwapien2012}, which states that, together with the acceleration of the information flow and processing across the financial markets, we can observe a significant change in the statistical properties of the returns at a particular time scale related to an effective acceleration of the market time with all of the possible consequences of this fact.

The results are mixed. On the one hand, the stock market indices (DJIA, DAX30, and S\&P500, for which the present results can be compared directly with earlier works) do not show any further signatures of the time acceleration compared with the data from 1998--1999 and 2004--2006. It seems that the acceleration that was reported \mbox{in~\cite{drozdz2003,drozdz2007}} stopped or was only a temporary effect. Such effects were already reported before for Asian markets~\cite{yang2006,eom2019} as well as in this work regarding the stocks, so they may be a source of the observed behavior. On the other hand, the results for the individual stock groups show that the market time acceleration can still be ongoing, but it is masked at the level of indices owing to the cross-correlations among the stocks that are now stronger and developing faster than even during the years 2004--2006~\cite{drozdz2007}. That particular time interval (2004--2006) was characterized by a volatility much smaller than in recent years, which witnessed large market events such as the flash crash on 5 February 2018, the coronavirus-related unsteadiness in early 2020 and the subsequent rally ending with new record highs of S\&P500 in August, the oil price drop in April 2020, etc. Large events, especially large falls, elevate the market correlation level, which can influence the statistical properties of data, including the distribution of returns. The auto- and cross-correlations are involved in an interesting interplay between two opposite-acting factors. The first factor is the market time flow speed, which works for market efficiency by shortening the period when the market autocorrelations are admissible. This factor shifts gradually the low-$\alpha$ behavior and the central limit theorem's realm to ever shorter time scales. The second factor is the asset cross-correlation strength, which causes thickening of the tails and decreases in $\alpha$ and $\beta$. It also violates the assumption of random variable independence and prevents the CLT from affecting the aggregated returns. This interplay and its consequences are interesting enough to be worthy of some more attention in future analyses. In particular, they can be responsible for the reported behavior of the return distributions in different time periods and suppressing the effects of the market time acceleration.

Currency exchange rates also no longer feel the market time acceleration such as that during 2004--2006~\cite{drozdz2007}, but now, not only is there no further time scale shortening but also a moderate step backwards is observed: the inverse cubic scaling is seen at longer time scales than in 2004--2006 but is still significantly shorter than that during the years 1987--1993~\cite{guillaume1997}. The cryptocurrencies (BTC and ETH) show the same crossover scale as before---equal to 1 h~\cite{drozdz2020b}. Since this market is relatively young, it underwent a phase of strong market time acceleration after 2013, and now, it seems to be stabilized. It is still the market that shows the most exemplary inverse cubic scaling behavior across different scales out of all the markets analyzed in this work. Gold price CFDs show a clear difference between the present results and the distribution tails over the years 1969--1999~\cite{matia2002} and 2012--2018~\cite{watorek2019} with increased tail slope during the recent years. In contrast, there is no  clear change in the tail slope regarding silver, high-grade copper, and crude oil.

It should be noted, however, that the CFD contract price quotations analyzed here are not precisely the same as the related asset spot price quotations, which the authors of other works dealt with. This difference may partially account for the difference in the outcomes. Finally, the COVID-19 pandemic outburst that took place in March--April 2020 in the U.S. constituted a strong perturbation to all the markets, caused large-amplitude price fluctuations, and led to a strong increase in the cross-correlations among many assets. For example, it resulted in decreasing distribution tail slopes for the CFD returns for crude oil and gold. Even more significant were the bitcoin fluctuations, which become L\'evy stable for the pandemic-outburst period.

In general, our results indicate that the monotonous shift in the time scales at which different types of dynamics can be observed in the financial data as well as the {related continuously} accelerating market time from past to present are oversimplified. In fact, there can be an underlying long-term trend of this type, but it is ``decorated'' with short-term phases of abrupt acceleration and, then, deceleration and stagnation. Our results indicate that the real market dynamics {consists} of continuous alternation of different regimes with different statistical properties that can form the overall impression of the market evolution direction. Together with the aforementioned problem of how the asset cross-correlations and the shortening autocorrelations compete against each other in shaping the statistical properties of data, it opens an intriguing direction for future work.

\newpage


\authorcontributions{Conceptualization, S.D., J.K., and M.W.; methodology, S.D., J.K., and M.W.; software, M.W.; validation, S.D., J.K., and M.W.; formal analysis, S.D., J.K., and M.W.; investigation, S.D., J.K., and M.W.; resources, M.W.; data curation, M.W.; writing---original draft preparation, J.K.; writing---review and editing, J.K. and M.W.; visualization, M.W.; supervision, S.D. All authors have read and agreed to the published version of the manuscript.}

\funding{{This research received no external funding.}}

\institutionalreview{{Not applicable.}}

\informedconsent{{Not applicable.}}

\dataavailability{Publicly available datasets were analyzed in this study. This data can be found at~\cite{dukascopy} and~\cite{kraken}.}


\conflictsofinterest{The authors declare no conflicts of interest.} 



%
\end{paracol}
\nointerlineskip

\appendixtitles{no} 
\appendixstart
\appendix
\section{}
\label{sect::app}


\vspace{-9pt}
\begin{specialtable}[H]\setlength{\tabcolsep}{2.9mm}
\widetable
\caption{List of stocks that are included in the stock sets considered in Section~\ref{sect::stocks}. Capitalization is given in billions ($10^9$) of currency units.}
\begin{tabular}{cccccccccccc}
\toprule
\multicolumn{2}{c}{\bf U.S. Large} & \multicolumn{2}{c}{\bf U.S. Medium} & \multicolumn{2}{c}{\bf U.S. Small} & \multicolumn{2}{c} {\bf U.K. Large} & \multicolumn{2}{c} {\bf German Large} & \multicolumn{2}{c} {\bf French Large} \\ \midrule
{\bf Ticker} & {\bf USD} & {\bf Ticker} & {\bf USD} & {\bf Ticker} & {\bf USD} & {\bf Ticker} & {\bf GBP} & {\bf Ticker} & {\bf EUR} & {\bf Ticker} & {\bf EUR} \\ \midrule
AAPL & 2000 & BIDU & 68.6 & CAG & 18.7 & RDSB & 107.4 & VOW3 & 140.6 & VK & 330.9 \\
MSFT & 1784 & NSC & 68.1 & MGM & 18.4 & ULVR & 105.7 & SAP & 126 & MC & 288 \\
AMZN & 1537 & CL & 67.7 & CAH & 18.3 & AZN & 94.2 & SIE & 112.5 & OR & 181.3 \\
GOOGL & 1378 & SHW & 67 & CTL & 18.1 & RIO & 88.7 & ALV & 89.5 & SAN & 105.2 \\
FB & 805 & SO & 65.9 & ULTA & 17.6 & HSBA & 86.4 & DTE & 81.8 & FP & 102.8 \\
BABA & 625 & APD & 63.1 & OMC & 16.3 & DGE & 70.4 & DAI & 80.6 & AIR & 78.9 \\
BRKB & 593 & ICE & 62.6 & TIF & 16 & BATS & 62.3 & BAS & 65.3 & KER & 74.8 \\
TSM & 592.8 & D & 60.9 & DVN & 14.8 & BP & 59 & MRK & 62.3 & SU & 73.8 \\
TSLA & 582 & ADSK & 59.5 & AAL & 14.7 & RB & 46.3 & DPW & 57.6 & AI & 65.9 \\
JPM & 473 & ADI & 56.6 & WYNN & 14.5 & BLT & 44.2 & BMW & 57.3 & BNP & 65.1 \\
V & 458 & ILMN & 56.2 & WHR & 14.1 & PRU & 40.6 & VNA & 56.7 & CS & 54.9 \\
JNJ & 438.5 & PGR & 55.9 & L & 13.8 & AAL & 39.5 & ADS & 52.8 & SAF & 51 \\
WMT & 383.5 & VRTX & 55.8 & SJM & 13.7 & GLEN & 38.1 & BAYN & 52.3 & DG & 50.8 \\
MA & 360 & BSX & 54.8 & TEVA & 12.7 & VOD & 37.7 & IFX & 47.6 & RI & 42.1 \\
UNH & 358.2 & EMR & 54.7 & IPG & 11.5 & REL & 35.5 & HEN3 & 38.5 & BN & 37.8 \\
DIS & 337 & NOC & 53.7 & DVA & 11.5 & LSE & 35.3 & MUV2 & 37 & ACA & 36.3 \\
PG & 336 & HUM & 53.3 & NWL & 11.4 & BARC & 31.7 & PAH3 & 28.7 & EDF & 35.4 \\
BAC & 331 & MET & 53.1 & GPS & 11.3 & NG & 30.7 & DB1 & 26 & VIV & 30.4 \\
HD & 324 & PBR & 53 & TAP & 11.3 & LLOY & 30.3 & EOAN & 26 & ENGI & 29.3 \\
NVDA & 318 & REGN & 51 & CF & 9.7 & CPG & 26.7 & RWE & 23.2 & ORA & 27.8 \\
PYPL & 277 & TWTR & 50 & NRG & 9.2 & CRH & 26.3 & CON & 22.8 & SGO & 26.8 \\
INTC & 261 & KHC & 49.9 & KSS & 9.1 & EXPN & 23.4 & DBK & 21.2 & CAP & 24.9 \\
CMCSA & 252.6 & NEM & 49.8 & MRO & 8.7 & RBS & 22.4 & FRE & 20.9 & LR & 21.3 \\
XOM & 243 & F & 49 & X & 6.9 & AHT & 20.1 & BEI & 20.5 & UG & 19.4 \\
VZ & 243 & DG & 48.6 & MAT & 6.9 & WOS & 20.1 & FME & 18.4 & GLE & 19.2 \\
KO & 231.6 & ITUB & 48.4 & APA & 6.8 & ABF & 19.4 & HEI & 15.3 & ALO & 16.2 \\
NFLX & 228 & MAR & 48 & AA & 6.1 & CCL & 18.1 & TKA & 7.1 & EN & 13 \\
ADBE & 224 & FCX & 47.7 & JWN & 6 & TSCO & 17.6 & CBK & 6.6 & PUB & 12.7 \\
CSCO & 222 & KMB & 47 & M & 5.1 & LGEN & 16.9 & LHA & 6.6 & VIE & 12.6 \\
T & 217.9 & LVS & 46.8 & EQT & 5.1 & ANTO & 16.7 & LXS & 5.5 & CA & 12.3 \\
 \bottomrule
\end{tabular}
\end{specialtable}




\reftitle{References}

\end{document}